\documentstyle[]{article}
\begin{document}

\title{Distribution of the magnetic field and current density
in superconducting films of finite thickness}

\author{D. Yu. Vodolazov, I.L.Maksimov 
\thanks{Corresponding author. Tel.: +7-831-2656255; Fax: +7-831-2658592;
E-mail: ilmaks@phys.unn.runnet.ru.} 
\thanks{Current address.  23 Gagarin ave., Nizhny Novgorod University, 
Nizhny Novgorod, 603600, Russia.} \\
Nizhny Novgorod University, 603600, Russia.}

\maketitle

\begin{abstract}
A one-dimensional equation describing the distribution of the effective 
vector potential $\overline A(y)$ across a film width, which holds for thin 
($d<\lambda$) and thick ($d>\lambda$) films alike, is derived based 
on the analysis of a 2D Maxwell-Londons equation for superconducting 
films in a perpendicular magnetic field. The validity of this equation 
for a finite-thickness film is verified by a numerical analysis. 
An approximation dependence $\overline A(y)$, finite 
(with all of its derivatives) across the entire film width,
is found for  films, being in the Meissner state. 
The flux-entry field is evaluated for a film of arbitrary thickness. 
An approximation expression is obtained for the distribution of the sheet 
current density in the mixed state of a pin-free superconducting film  
with an edge barrier. The latter approximation allows to estimate 
 magnetic field concentration factor at the film edge as a function  of
 external magnetic field  and geometrical parameters of the sample.

\end{abstract}



\section{Introduction}

Of active interest recently is theoretical \cite{1,2,3,4,5,6,7,8,9} 
and experimental \cite{9,10} investigation of mixed static and dynamic 
states in superconducting films in a perpendicular magnetic 
field. Theoretical calculations of various magnetic characteristics 
of films in such a geometry, using the microscopic theory or the 
Ginzburg-Landau equations does not seem possible because of the 
mathematical intricacies involved in their solution.Therefore, 
in practical calculations the following two approaches are mostly 
employed. In the first method the Maxwell equation is analysed 
jointly with the London equation, which yields an integral-differential 
equation \cite{1,4,5,6,7,8} for distribution of the current density 
integrated over the film thickness $d$. This equation was derived on the 
assumption that the film is thin ($d \ll \lambda$, $\lambda$ is the 
London penetration depth), which naturally limits its applicability range.

The other approach is based on the theory of complex variable functions 
(TCVF) \cite{3,9}, used in transformation of integral equations \cite{2}. 
In this case the phenomenological dependences ${\bf B(H)}$ and 
${\bf J(E)}$ are employed as an additional condition instead of the 
London equation. It should be noted that neither of these methods has 
ever focussed on investigating the effects related to finiteness of the 
London penetration depth $\lambda$. Besides, the equality of the results 
obtained through solving of the integral equation for thin films and by 
the use of the TCVF methods for thick films leads us to believe that for 
films of arbitrary thickness there should exist one equation 
describing distribution of the current density and the magnetic 
field across the film width.

The present paper deals with a study on the distribution of the magnetic 
field and current density in the Meissner and mixed states for films 
placed in a perpendicular magnetic field. It is shown that for the 
finite-thickness films the Maxwell-London equation \cite{1,4} describes 
distribution of the vector potential (current density) averaged 
over the film thickness $d$, provided the latter, is much smaller 
than the film width $W$: $d/W\ll 1$. For thin films, $d \ll \lambda$, 
this equation is valid practically over the entire width of 
the film, while in the case $d \gg \lambda$ it holds everywhere 
in the film, except for the areas near the edges, which measure 
as $W/2< |Y| \leq W/2-d/2$.

The paper is organized as follows. Section 2 describes derivation 
of an one-dimensional equaton for distribution of the 
film-thickness-averaged vector potential across a sample width, 
based on the analysis of 2D Maxwell-London equation. In subsections
2.1 and 2.2 the 2D and 1D equations are numerically studied and 
compared for thin and thick films respectively. An approximation 
dependence for $\overline A(y)$ is obtained through numerical 
solution of these equations. The distribution of the vector potential 
(or the local current density) and of the local magneti field over a 
superconductor cross-section is described. Section 3 deals with 
estimation of the field for the first vortex entry into thin and 
thick superconducting films (subsection 3.1 and 3.2 respectivly). 
In subsection 3.3 we discuss the influence 
of surface defects and a layered structure of superconductors on a 
barrier suppression field value. Section 4 is concerned with the 
structure of a mixed state in thin and thick superconducting films of an 
arbitrary width in the absence of bulk pinning. An approximation 
formula is found for the integral (over thickness) current density, 
which is then used as a basis for constructing the magnetization curves 
for these superconductors. Section 5 sums up the main results obtained 
in this work.

\section{The structure of the Meissner state}

Assume an infinite (in the $X$ direction) superconducting film of 
width $W$ and thickness $d$ in a perpendicular magnetic field; the 
geometry is shown in Fig.1. Let us first consider the Meissner state. 
The Maxwell equation has the form (in a gauge $\nabla \cdot {\bf A}=0$)

$$
\Delta {\bf A}=-\frac{4\pi}{c} {\bf j}, \eqno(1)
$$

where $\Delta$ is the 2D Laplacian operator. It also follows from 
the symmetry of the problem that only the $x$ components of the vector 
potential ${\bf A}=(A_x,0,0)$ and of the current density 
${\bf j}=(j_x,0,0)$ are not zero. The boundary conditions are 
$\frac{\partial A_x}{\partial Y}|_{Y \to \pm \infty}=-H_{\infty}$,  
$\frac{\partial A_x}{\partial Z}|_{Z \to \pm \infty}=0$, where 
$H_{\infty}$ is the field far from film ${\bf H}=(0,0,H_{\infty})$. 
It should be noted that by 
the magnetic field $H$ in a film here we imply a microscopic field 
averaged over scales much larger than the atomic one but much 
smaller than $\lambda$. 

In this case it is convenient to change over from the differential 
equation (1) to its integral analog. Using the Green function of the 2D 
Laplacian operator, we rewrite Eq.(1) as

$$
A_x(Y,Z)=A_x^0(Y)-\frac {2}{c} \int_{-W/2}^{W/2} 
\int_{-d/2}^{d/2} (\ln|{\bf R} - {\bf R'}|+C)j_x(Y',Z')dY'dZ', \eqno(2)
$$    

where $C$ is the constant generic for the 2D Green function, 
$A_x^0(Y)$ is the vector potential of an external field: 
$A_x^0(Y)=-H_{\infty}Y$ and the origin of coordinates 
is chosen in the film centre.

Employing  London equation ${\bf j}=-á{\bf A}/4\pi\lambda^2$ 
and introducing dimensionless coordinates $y=2Y/W$, $z=2Z/d$,
 Eq.(2) reads

$$
\displaylines{A_x(y,z)=A_x^0(y)+\frac {Wd}{16\pi \lambda^2} 
\int_{-1}^{1}\int_{-1}^{1} \ln\left((y-y')^2+\left(\frac{d}{W}\right)^2 
(z-z')^2\right)A_x(y',z')dy'dz'\cr
\hfill +\frac{Wd}{16\pi\lambda^2} \tilde C \int_{-1}^{1} 
\int_{-1}^{1} A_x(y',z')dy'dz',\hfill\llap{(3)}\cr}
$$

where $\tilde C=C+2\ln(W/2)$. 
The latter integral in (3) is directly proportional to the total 
current. In a magnetic field (without a transport current) the 
total current is equal to zero due to the symmetry, so the last term
in the right-hand side of Eq. (3) vanishes. We now 
average Eq.(3) over the film thickness, which yields the 
following expression (hereafter by $\overline A(y)$ we 
imply 0.5$\int_{-1}^{1}A_x(y,z)dz$)

$$
\displaylines{\overline A(y)=-H_{\infty}yW/2+\frac {Wd}{4\pi \lambda^2} 
\int_{-1}^{1} \ln|y-y'| \overline A(y')dy' \cr
\hfill +\frac {Wd}{32\pi \lambda^2} \int_{-1}^{1} \int_{-1}^{1} \int_{-1}^{1}
\ln(1+b^2/(y-y')^2)A_x(y',z')dy'dz dz',\hfill\llap{(4)}\cr}
$$    

where the integral kernel of Eq.(3) is written in the form

$$
\ln\left((y-y')^2+\left(\frac{d}{W}\right)^2(z-z')^2\right)
=2\ln|y-y'|+\ln\left(1+\frac{b^2}{(y-y')^2}\right)  
$$

and the designation $b=(z-z')d/W \ll 1$ is introduced (as in this 
case of superconducting films $d/W \ll 1$).

The function $\ln(1+b^2/(y-y')^2)$ is not zero only in a small 
region around point $y'$: $|y-y'|\leq |b|$. In this case integration 
over $y'$ in the second integral of Eq.(4) can be done only over  
this small region. For the same reason we can expand the function 
$A_x(y',z')$ into the Taylor series in terms of $y'$ near point $y$:

$$
A_x(y',z')=A_x(y,z')+\frac{\partial A_x(y,z')}{\partial y}(y'-y)+ . 
\eqno(5)
$$

Note that expansion (5) (in the above specified limit) is valid for 
thin ($d<\lambda$) films over the entire sample width.
In thick ($d>\lambda$) films the validity of this expansion is 
violated in the near-edge regions with dimensions of order $d/W$. 
More details on the applicability of Eq.(5) are provided in the end 
of this Section.

After the series-expansion of function $A_x(y',z')$ (5) it is now 
possible to calculate the last term of Eq.(4) 

$$
\int_{y-b}^{y+b}\int_{-1}^{1}\int_{-1}^{1} \ln \left(1+
\left(\frac{d}{W}\right)^2(z-z')^2/(y-y')^2\right)
\left(A_x(y,z')+\frac{\partial A_x(y,z')}{\partial y}(y'-y)+...\right)
dy' dz dz', \eqno(6)
$$    

\noindent
(note, that the upper (lower) limit of the integration over $y'$ will 
change by $1(-1)$, when point $y$ becomes close to the film edge, i.e., 
when $1-|b|\leq |ã|<1 $; this takes place because the integration in (4) 
is carried out only over a sample volume) and to show that (6) is 
equal to zero in a wide parameters range.
Indeed, integration of Eq.(6), first in terms of $y'$ and then in $z$ 
and $z'$, provides a direct evidence (bearing in mind that function 
$A_x(y',z')$ is even in $z'$) that integral (6) is zero 
for all values of $y$ satisfying the inequality $|y|\leq 1-|b|$. 
In the region $1<|y|\leq 1-|b|$ integral (6) leads to appearance 
of nonzero terms that for thin films are small due to the presence of 
the corrections of $(d/\lambda)^n$ ($n>1$) type. They have to be taken into 
account when we are interested in, for example, the distribution of 
the derivative $d\overline A/dy$ near the film edges (since disregard 
for these terms will cause a logarithmic divergence of the first 
derivative). For thick films, allowance for the near-edge regions of a 
superconductor in integral (6) cannot largely affect the $\overline A(y)$ 
distribution off the film edges because of smallness of $|b|$.

Thus, the 2D equation (3) is reduced to a 1D equation for the 
film-thickness-averaged vector potential $\overline A(y)$, which is 
valid in the region $|y|\leq 1-|b|$

$$
\overline A(y)=-H_{\infty}yW/2+\frac {Wd}{4\pi \lambda^2} \int_{-1}^{1}  
\ln|y-y'|\overline A(y')dy' \eqno(7)
$$    

In the following sections the results of a numerical solution of 
Eqs. (7) and (3) are provided for thin and thick films. 

\subsection{Thin films ($d<\lambda$)}

It turns out that in films satisfying the condition 
$d/\lambda \leq 1/4$ the difference between the solutions of Eqs (7) 
and (3) (averaged over thickness) is about $1 \%$ far from film edge 
and less than $4 \%$ in a narrow near-edge region. An 
appreciable error in the near-edge region (which is slightly growing towards 
the film edge with a larger numerical step) depends on the presence 
of the small corrections in (7) that were ignored. 

Besides, the vector potential in this case is practically independent 
of $z$ (but not the derivative $\partial A_x(y,z)/\partial z$). Thus, 
at $d=\lambda$ the relation $A_x(y,1)/A_x(y,0) \approx 1.07$, i.e., 
the difference is about $7 \%$. With a lower $d/\lambda$ ratio 
the relation $A_x(y,1)/A_x(y,0)$ tends to unit. 

We have derived an asymptotic expression for the vector potential 
distribution, satisfying Eq.(7) (and, hence, (3) averaged over $z$) 
with a sufficently high accuracy (not less than $3 \%$ at a film edge and 
far from edge region, and not less than $6 \%$ in the near-edge 
region, see Fig.2):

$$
\overline A(y)=-\frac{\lambda_{eff}H_{\infty}y}{\sqrt{\alpha(1-y^2)+
\beta}}, \eqno(8)
$$

where $\lambda_{eff.}=\lambda^2/d$, $\beta \simeq 2\lambda_{eff}/\pi W+
4(\lambda_{eff}/W)^2$, and the dependence $\alpha(W/\lambda_{eff})$ 
is shown in Fig.3 together with the approximation (9)

$$
\alpha \approx 0.25-\frac{0.63}{(W/\lambda_{eff})^{0.5}}+\frac{1.2}{
(W/\lambda_{eff})^{0.8}}. \eqno(9)
$$

At $W<\lambda_{eff}$ the dependence $\overline A(y)$ becomes almost linear. 
Formula (8) with $\alpha=0.25$, $\beta=0$ was derived analytically in 
\cite{1} by solving Eq.(7) (to be more exact, a simplified version of
Eq.(7) in which the left-hand part is omitted, which corresponds to the 
condition $W\lambda_{eff} \gg 1$).

The resultant dependence $\overline A(y)$ allows to calculate the 
field concentration parameter $\gamma=\overline H_z^{edge}/H_{\infty}$ 
(where $\overline H_z^{edge}$ is the edge field averaged over 
superconductor thickness) for films of such type:

$$
\gamma=\frac{\overline H_z^{edge}}{H_{\infty}}=\frac{2\lambda_{eff}}
{W\sqrt{\beta}}\left(\sqrt{1+\frac{\alpha}{\beta}} \right) \eqno(10)
$$

At $W \gg \lambda_{eff}$ Eq.(10) transforms into

$$
\gamma=\frac{\pi\sqrt{2\pi}}{10}\sqrt{\frac{W}{\lambda_{eff}}},  
$$

where the coefficient preceding $\sqrt{W/\lambda_{eff}}$ is a quantity 
of order unit. The difference between Eq.(10) and the numerically 
obtained expression for $\gamma$ may reach $30 \%$: for wide films, 
($W \gg \lambda_{eff}$), Eq.(10) yields an overestimated result, see 
insert in Fig.4. This takes place because, unlike the $\overline A(y)$ 
function itself, the first derivative (8) with respect to $y$ (magnetic 
field) adequately satisfies the numerical solution everywhere except for 
the narrow region near a film edge (see Fig.4). Due to the logarithmic 
divergence of the magnetic field at a film edge, which follows from 
the solution of Eq.(7), the approximation expression for 
$\overline H_z^{edge}$ was compared with the numerical solution of Eq.(3). 
Fig.4 also shows the interpolation function $1+0.66\sqrt{W/\lambda_{eff}}$ 
for the numerical analysis data (see solid line in the insert). Thus, 
the difference between (10) and the numerical result in the wide film limit 
$W/\lambda_{eff} \gg 1$ comes to about $17 \%$.

\subsection{Strip of finite thickness ($d\gg \lambda$)}

A comparative numerical analysis of the solutions to Eqs (3) integrated 
over a superconductor thickness and (7) 
was also carried out for the case $\lambda \ll d \ll W$. 
It was found out that the solutions coincide (to the 
accuracy of about $3 \%$) in the region $|y|<1-d/W$ and (
quite surprisingly) in points $|y|= 1$. In the near-edge region 
we observed a difference in the solutions of Eqs.(3) and (7). 

An approximation equation has also been derived for the dependence 
$\overline A(y)$. It turned out to be exactly the same as 
the dependence (8) with the selection $\alpha\approx 0.25$, 
$\beta \approx 0.64\lambda^2/dW$. One can easily see that the obtained 
values for $\alpha$ and $\beta$ practically coincide, to a numerical 
error, with those obtained for thin films at $\lambda^2/Wd \ll 1$. 
Fig.5 shows the dependence $\overline A(y)$ derived from the solution 
to equation (3), and also Eq.(8). It is seen from the latter that 
maximum deviation of the solution to Eq.(3) from (8) (and, hence, 
from the solution to (7)) occurs in the region $y>1-d/W$, but in 
point $y=1$, however, both solutions coincide again. 
Thus, expression (8) provides an adequate description of the 
$\overline A(y)$ distribution (the difference from the numerical 
solution of Eq.(3) does not exceed $3 \%$) in the region 
$|y|\leq 1-d/W$ and in points $|y|=1$. This confirms the above 
statement that Eq.(7) describes well the distribution of the 
$\overline A(y)$ dependence in a film depth. Besides, Eq.(7) is 
apparently valid immediately at the edge of a superconductor as well,
which, in our opinion, is quite surprising fact. 

Fig.6 shows the distribution of the current density over a 
film cross-section and the distribution of the absolute value of 
local magnetic field ($|H|=\sqrt{H_z^2+H_y^2}$) both inside and outside a 
film with dimensions $W=100\lambda$, $d=10\lambda$. As seen from 
Fig.6a,c, the magnetic field reaches its maximum at the corners 
(side edges) of a superconductor. At the equator $(y=1,z=0)$ 
the magnetic field is less intensive, but not appreciably smaller than 
the field at the corners (for comparison, at the corners 
$|H|\approx 4.1H_{\infty}$, while in the middle of a side face 
$|H|\approx 2.9H_{\infty}$ for the given parameters of film). 
It is easily seen that towards the film interior (along the $y$-axis) the 
magnetic field is practically uniform through the entire sample thickness,
except for the near-surface areas with the thickness of order $\lambda$
where magnetic lines abruptly change direction. A similar behaviour is 
demonstrated by a current density (see figs.6 b,d). It is proved 
numerically that both the current density and the magnetic field 
fall off towards a sample centre by a law similar to the 
exponential one, not only from the side faces but also from the top 
and bottom ones. So, in thick films $\lambda \ll d \ll W$ 
screening currents flow only in the near-surface layers of 
thickness about $\lambda$. On the same scale there is a decrease 
of a local magnetic field in superconducting samples of such type.

The numerical solution of Eq.(3) also provides a possibility to 
determine the field at a film edge. Unfortunately, unlike with 
thin films, Eq.(8) differs largely from the numerical result for 
the near-edge region (this discrepancy may reach $30\%$). Therefore, 
the use of (8) in calculations of a thickness-averaged $z$-component of the 
magnetic field at a film edge certainly leads to a considerable error. 

In Fig.7 the obtained numerical dependences $\overline H_z^{edge}/H_{\infty}$
(the $z$-component of magnetic field, averaged over thickness) and
$H_z(1,0)/H_{\infty}$ (the $z$-component of magnetic field on equator)
on $\sqrt{W/d}$ are presented. It is clearly seen that with a good 
accuracy the dependences are linear even for the $W/d$ values which are 
close to unit. Note that the coefficient of proportionality between 
$\overline H_z^{edge}/H_{\infty}$ ¨ $\sqrt{W/d}$ is equal to $\sim 1.03$, 
which practically is the same as its estimate ($\sim 1$) found in \cite{9}. 
Generally speaking, the value of this coefficient depends on a film 
thickness (or, rather, the ratio $d/\lambda$). Thus, the insert in Fig.7 
illustrates relationships $\overline H_z^{edge}/H_{\infty}$, 
$H_z(1,0)/H_{\infty}$ for various values of film thickness and 
shape parameter $W/d=5$. It is seen that only the quantity 
$H_z(1,0)$ is practically independent of the ratio $d/\lambda$. 
The strongest dependence on film thickness is exhibited by $H_z(1,1)$ 
(it grows with the increase of the ratio $d/\lambda$). This results 
in a slight increase of the average field $\overline H_z^{edge}$ with a 
growth of film thickness (given the same $W/d$ ratio). However, for 
very thick films, $d \gg \lambda$, $\overline H_z^{edge}/H_{\infty}$ 
is supposed to practically cease to be dependent on $d/\lambda$. 
Indeed, in the limit of interest the equator field is independent 
of $d/\lambda$, while the corner field $H_z(1,1)$ increases as $\sqrt{W/\lambda}$ 
(which was derived from the expression (11c,d) given below)). The sharpest 
variation of the magnetic field intensity occurs at a distance of order 
$\lambda$ from the top/bottom surfaces. Correspondingly, the contribution 
of those regions in $\overline H_z^{edge}/H_{\infty}$  will be of 
order $\lambda\sqrt{W/\lambda}/d=\sqrt{W/d}\sqrt{\lambda/d}$ and 
will become negligible with the increase of the ratio $d/\lambda$.

Besides the approximation expression for $\overline A(y)$, we have 
found numerical estimates for the vector potential in points 
($1,1$) (on a corner); ($1,0$) (on the equator), and also 
the distribution of the vector potential (current density) over the 
upper/lower surfaces:

$$
A_x(1,0) \simeq -H\sqrt{\frac{W}{d}}\lambda, \eqno(11a)
$$

$$
A_x(1,1) \simeq A_x(W/2,0)\left(1+\frac{1}{\sqrt{16\pi}}\frac{d}{\lambda}
\right)^{1/2}, \eqno(11b)
$$  

$$
A_x(y,\pm 1)=-\frac{\lambda_{eff}H_{\infty}y}{\sqrt{\alpha(
1-y^2)+\beta}}, \eqno(11c) 
$$

where

$$
\alpha \simeq \frac{0.25}{(1+(d/\lambda)^2/2\pi)}, \, \, \,  
\beta \simeq \frac{2\lambda_{eff}}{\pi W(1+(d/\lambda)/\sqrt{16\pi})}. 
\eqno(11d)
$$

It is easily seen that at $d\gg \lambda$ the vector potential in 
points $(\pm 1, \pm 1)$ will largely exceed its value in points 
$(\pm 1,0)$. Using expression (11c) which is similar to 
(8) with renormalized parameters $\alpha$, $\beta$, we can find 
the points (lines) on the upper and lower surfaces, 
where the vector potential will coincide in absolute value with 
that on the equator. Simple calculations show that these lines 
are at a distance $\sim d/\pi$ from the side surfaces 
of the strip having $\lambda \ll d \ll W$.

These results allow to confirm the assumption (see subsection 2.1) on the
possibility of expanding $A_x(y',z')$ in the limits $(y-b,y+b)$.
Indeed, in the case of thin films the vector potential (or current
density in a mixed state; see Sec. 4) varies on scales much larger 
than $\lambda$ and, hence, than $d$ (see (8)). For thick films, the 
vector potential (current density) far from edges is finite only in 
the surface layers of thickness of order $\lambda$. At the same time, 
the scale of variation for $A_x(y,z)$ along $y$ off sample edges is 
macroscopic (see expressions (11)). Therefore, expansion (6) is also 
possible off the edges. Near the edges, however, $A_x(y,z) \neq 0$ over 
entire thickness, and the scale of $A(y,z)$ variation (see fig.6b,d) 
in this region is $\lambda$ (in the $y$ direction).
Hence, expansion (6) is not valid in the limits $y-b\leq y'\leq y+b$ 
near the edges ($|y| \to 1$) of thick superconducting film.

\section{The conditions for vortex entry in superconducting films}

Using expressions (8,11) it is possible to estimate the edge barrier 
suppression field $_s$ or the first-vortex entry field  into 
superconducting films. 

\subsection{Thin film}

The vortices may enter into a thin film in the Meissner state 
provided the condition $|\overline A(\pm 1)|=A_{crit}$ is met; 
here $A_{crit}\approx \Phi_0/2\pi\xi$ \cite{11,12} ($\Phi_0$ 
is the quantum of a magnetic flux, $\xi$ is the coherence length). 
The resultant expression for $H_s$ is

$$
H_s \approx \frac{\Phi_0}{2\pi\xi\lambda_{eff}}\sqrt{\beta},
\eqno(12)
$$   

with $\beta$ being the same as in the expression (8).
Dependence (12) in the limit $W \gg \lambda_{eff}$ and 
$W \ll \lambda_{eff}$ coincides to a factor of order one 
with the expression for the Meissner state breakdown field obtained 
in the limiting cases of wide and narrow thin films in \cite{1}. From (10,12) 
we can easily find the value of the magnetic field (or, rather, the 
thickness-averaged z-component) at a film edge $\overline H_z^{edge}$, 
when vortices start penetrating in it. For example, for wide films 
$\overline H_z^{edge}$ is

$$
\bar H_z^{edge} \approx \frac{\Phi_0}{10\xi\lambda_{eff.}}.
$$ 

To an accuracy of the factor of order one the above expression coincides 
with the field in the core of a Pearl vortex which is equal 
to $\Phi_0/4\pi\xi\lambda_{eff}$ \cite{13}.

\subsection{Thick film}

The main difference between thick and thin films is that the vector 
potential for the former largely depends on $z$. However, we should 
apparently anticipate that the conditions of vortex entry here will be 
qualitatively similar to those for thin films. Indeed, as shown in 
\cite{11,12}, after the vector potential has reached its 
critical value at a superconductor edge in the Meissner state 
(in the mixed state it is the gauge-invariant potential 
${ \bf \Pi}=\Phi_0\nabla \varphi/2\pi - {\bf A}$ that should reach a 
critical value), the order parameter $\Psi=|\Psi|e^{i\varphi}$ is 
strongly suppressed and vortex formation begins. The above papers 
dealt with bulk superconductors and thin-film samples, which, due to 
the symmetry of the problem or problem geometry, could be assumed 
homogeneous along the $z$-axis.

It should be expected that in thick films the order parameter will 
be suppressed in the regions where the vector potential $A_x(y,z)$ 
reaches its critical value.

First, the condition $|A_x(y,z)|=A_{crit}$ will be satisfied at the 
side edges $(\pm 1, \pm 1)$ of a superconductor 
(as the magnetic field is increasing 
from zero). It means that the order parameter will be suppressed in 
these points first. With a further increase of the magnetic field 
this situation may develop by two scenarios:

1. Suppression of the order parameter results in tilted 
vortices that start to form at the {\it corners} of a superconductor
cross-section. When the vector potential at the {\it equator} reaches the 
critical value, 
two tilted vortices fragments (from top and from bottom) will join
each other to form one rectilinear vortex. In the absence of pinning 
the latter is able to penetrate into the sample centre
driven by the Lorentz force. 
Similar vortex entry scheme was discussed in \cite{9}.

2. In the course of further magnetic field 
increase, the order parameter becomes suppressed in the 
region of side edges. This causes areas with a suppressed order parameter 
to appear near the side edges, which would allow to regard a film 
cross-section as a rectangular with rounded-off edges. It should be 
emphasized that the geometrical sizes of sample remain unchanged in this 
situation, only its physical properties vary in the regions near side 
edges. This scenario allows to explain the physical mechanism behind 
the formation of the "geometrical rounds-off" near the corners 
of a rectangular cross-section sample, which were 
considered in \cite{3}. However, unlike \cite{3,9}, our approach is 
based on the assumption that vortices will start entering deep inside
a superconductor 
when the condition $|A_x(y,z)|=A_{crit}$ is fulfilled at the {\it equator}. 
By that moment the effective "round-off" radius will reach a value of 
order $d/2$.

Which of these two scenarios is practically feasible can be found out 
only through experiment. Theoretically this question can be answered 
by numerical solution of a problem on a vortex entry in a 3D sample 
within the nonstationary theory of Ginzburg-Landau.

The feature common for the above two schemes is, actually, the assumption 
that vortices enter deep a thick film when the condition 
$|A_x(1,0)|=A_{crit}$
is met. This allows to estimate the field of vortex entry inside a sample. 
Using Eq.(11a) (regardless of possible variation due to the penetration of 
tilted vortices or the existence of areas with a suppressed order 
parameter), we now derive the expression for field $H_s$, which is 
equivalent (12) (with $\beta=2\lambda_{eff}/\pi W$). 
This similarity is due to the fact that the $A_x(\pm 1,0)$ 
value is defined practically by the same expression for both thin 
and thick films, provided parameter $\lambda^2/dW$ is the same.

By analogy with thin films, one can find a value of the
field at a film edge when the vortices start to penetrate deep into a sample. 
Yet, as opposed to thin films, field 
$H_z^{edge}$ largely depends on the $z$ coordinate in the 
sample region. Estimations of the equator field yields

$$
H_z^{eq}=H_z(1,0)\approx \frac{\Phi_0}{2\pi\xi\lambda},
$$  

which is practically equivalent to the $\overline H_z^{edge}$ (see fig.7).
Note also that $H_z^{eq}(H_s)$ to the factor of order unit is equal to 
the thermodynamical field $H_c$, or the surface barrier suppression 
field for bulk superconductors in the absence of edge defects.

\subsection{The influence of surface defects and anisotropy}

The resultant values for field $H_s$ are characteristics of isotropic 
superconductors with ideal surface. As was established in 
\cite{12,14}, surface defects can considerably decrease the value of $H_s$. 
For example, in \cite{14} for the case $\kappa \gg 1$ ($\kappa$ 
is the Ginzburg-Landau parameter) maximum suppression of the entry field 
$g=H_s/H_{en}$ (where $H_{en}$ is the field of vortices entry in a 
superconductor with surface defects) was estimated as 
$g \approx \sqrt{\kappa/\pi}$. Thus, for $\kappa=100$ we will have 
$g \simeq 5.6$.

A strong influence  on the value of $H_s$ may be produced also by 
anisotropy or, rather, layered 
structure of superconductors. If the layers are not Josephson-coupled 
(or are weakly coupled), a superconductor should be regarded as a stack 
of superconducting layers. 
This geometry can be simulated, if we multiply the integrand in equation 
(3) (and, hence, in integral (6)) by the step periodic function $z$ 
which is equal to one in the superconducting layer and is zero 
in the interlayer space.

Let a period in such a structure be much smaller than $\lambda$ 
(which is practically always fulfilled for $HTSC$), a layer thickness 
be $l$ and an interlayer separation be $m$. Then we can assume the 
distribution of $A_x(z)$ to be a smooth function $z$, similar to the 
dependence $A_x(z)$ for a homogeneous superconductor. In this case, 
the integral in Eq.(3) for a layered superconductor will be $(l+m)/l$ 
times smaller than that for an isotropic superconductor. In other words, 
we can replace this integral for a layered superconductor by the integral 
for an isotropic case by introducing an effective penetration depth
$\lambda'=\lambda \sqrt{(l+m)/l}$. 
In this way we immediately obtain the distributions of $\overline A(y)$, 
$A_x(y,1)$, and also the values for $A_x(y,z)$ at the side edges 
and the equator with allowance for the layered structure of superconductor.

One particular effect of the anisotropy is that the value 
of $A_x(1,0)$ will be $\sqrt{(l+m)/l}$ times larger than that for an 
isotropic superconductor, all other conditions being equal. 
This, in its turn, will lead to a $\sqrt{(l+m)/l}$ times smaller field of 
vortex entry in a superconductor. For example, at $l=3\dot A$ and 
$m=12\dot A$, typical for $BiBaCuO$, one finds $\sqrt{(l+m)/l}\approx 2.2$.

Thus, the two above-mentioned factors, i.e., surface defects and layered 
structure of superconductors may cause a considerable (10-fold and more) 
decrease of the vortex entry field in layered superconductors with 
surface defects, as compared to the vortex entry field in isotropic 
perfect-surface superconducting films. 

Another conclusion following from the fact of a layered structure in such 
systems is that the scale of a local magnetic field penetration, for example, for thick 
films, will be $\lambda'=\lambda\sqrt{(l+m)/l}$, and for thin films 
the parameter $\lambda_{eff}$ has to be changed for $\lambda'_{eff}=
\lambda'\,^2/d$.
At the same time, the thickness-averaged current density, unlike the 
thickness-averaged vector potential, will practically remain unchanged 
across the entire film width, except for the regions lying close to 
sample edges. This can be accounted for by the fact that the expression 
for current density $j(y)=-cA(y)/4\pi \lambda^2$ will include $\lambda$ 
only at a superconductor edge (see (8)). Likewise, all other quantities that 
obviously do not include $\lambda$ (for example, a degree of the 
magnetic field concentration at the equator of thick film) will 
remain the same.

\section{The structure of a mixed state}

Let us now discuss the parameters of a mixed state arising in a 
superconducting film in fields larger than $H_s$. Here we neglect the 
presence of bulk pinning, which is justified for soft 
superconductors. This problem was studied earlier in \cite{3,5,6,7,9,15}. 
In \cite{7,15} the authors considered a case of narrow thin films 
$Wd/\lambda^2 \ll 1$, \cite{5} deals with wide thin films $Wd/\lambda^2\gg 1$, 
and \cite{3,9} is a study on thick films that formally obey the 
condition $Wd/\lambda^2 \gg 1$. We will analyse a general case to show that 
it embraces either of the above two situations. Besides, the resultant 
distribution of current density will be finite across the entire film 
width, as opposed to the results of the above cited works.

Consider a superconducting film in a mixed state, placed in a perpendicular 
magnetic field. The current density and the vector potential in the 
London limit are related as 
${\bf j}=-({\bf A}-\Phi_0\nabla \varphi/2\pi)c/4\pi \lambda^2$, 
where $\varphi$ is the order parameter phase. The Maxwell 
equation will have the form (1), in which $\Delta$ is now a 3D 
Laplacian operator. Using the Green function for $\Delta$, we can write 
this equation in an integral form:

$$
{\bf A}({\bf r})={\bf A}^0({\bf r})+\frac {1}{c} \int \int 
\int \frac{{\bf j}({\bf r'})}{|\bf r-r'|}dx'dy'dz'. \eqno(13)
$$    

We now subtract function 
$\nabla \varphi({\bf r})$ from the left- and right-hand parts of Eq.(13) 
and take curl from these parts (using the property
$\nabla \times \nabla \varphi({\bf r}) = 
2\pi \delta({\bf r}- {\bf r}')$, where ${\bf r}'=(x',y')$ are the 
vortex coordinates). Next, we do the averaging over coordinates $x,y$ 
on scales much larger than an intervortex distance. After these 
operations, the distribution of a current density becomes uniform 
along the $x$-axis and we can perform integration over $x'$ in (13). 
Then we average the obtained equation over a film thickness and use 
the results of the integral (6) calculations. This will yields an equation 
for the sheet current density $i(y)=\int_{-d/2}^{d/2}j_x(y,z)dz$, 

$$
\frac{8\pi\lambda_{eff}}{cW}\frac{di(y)}{dy}+\frac{2}{c}
\int_{-1}^{1} \frac{i(y')}{y-y'}dy'=-H_{\infty}+
n(y)\Phi_0, \eqno(14)
$$

where $n(y)$ is the density of vortices, the distance being measured 
in units of $W/2$. For the first time this equation was derived in \cite{4} 
for thin films $d \ll \lambda$. Just like Eq.(7), (14) is valid at 
$|y| \leq 1-|b|$ for thick films, and across the width for thin 
films (excluding an extremely narrow near-edge region of size 
$d \ll \lambda$). Besides, it should be 
expected by analogy with the Meissner state that Eq.(14) for thick 
films will also be valid directly at a sample edge.

We analysed Eq.(14) numerically for different values of parameter 
$W/\lambda_{eff}$, using the condition that current density is zero 
in the region where vortices exist, and takes finite value in vortex-free 
regions \cite{3,5,6,9}. Besides, we set a boundary condition 
$j(\pm 1)=\pm j_s$ on a current density (in increasing magnetic field), 
which allows for an edge barrier. The value of the 
current density $j_s$ of order of the Ginzburg-Landau depairing 
current density for ideal-surface superconductors \cite{11,12}. 
In result, we have obtained the approximation expression for $i(y)$

$$
i(y)=
\left \{ \begin{array}{ll}
\displaystyle{0} & 0<|y|<a ,\\ 
\displaystyle{\frac{cH_{\infty}(z+1)}{4\pi \sqrt{\alpha(1-z^2)+\beta\frac{(|y|+a)^2}
{(1+a)^2}}} {\rm sign}(y)} & a<|y|<1 ,
\end{array} \right. \eqno(15)
$$

where

$$
z=\frac{2(y^2-a^2)}{(1-a^2)}-1,
$$

$$
\beta \simeq \frac{8}{\pi} \frac{1}{1-a^2} \frac{\lambda_{eff}}{W}+
\frac{16}{(1-a)^2} \left( \frac{\lambda_{eff}}{W} \right)^2,
$$

$a(H_{\infty})$ is the half-width of the vortex-filled region
and parameter $\alpha$ is defined by expression (9) in which W has been 
replaced by $W(1-a)$.

Expressions (15), being not derived, represents a rather useful 
interpolation for the distribution of the sheet current density 
for a film in a mixed state. We would like to emphasize again that in the 
thin-film case $W/\lambda_{eff}$ can be both smaller (narrow films) 
and larger (wide films) than unit, whereas for 
thick films this ratio is always much larger than 1. 

Fig.8 shows the dependence $i(y)$ for different values of a magnetic 
field. The difference of approximation (15) from the numerical result 
does not exceed $4 \%$ in the vortex-free zone ($a<y<1$). Note, that
in the near-edge region and close to the boundary of the 
vortex-filled region deviation may come to about $9 \%$.

The dependence $a(H_{\infty})$ (in increasing magnetic field) 
is to be found from the following expression:

$$
\frac{8}{\pi}\frac{1}{1-a^2} \frac{\lambda_{eff}}{W}+\frac{16}{(1-a)^2}
\left( \frac{\lambda_{eff}}{W} \right)^2= \left( \frac{H_{\infty}}{H_s}
\right)^2 \left(\frac{8}{\pi} \frac{\lambda_{eff}}{W}+
16 \left(\frac{\lambda_{eff}}{W}\right)^2 \right).
$$

For $W/\lambda_{eff}\gg 1$ we have

$$
a(H_{\infty})=\sqrt{1-(H_s/H_{\infty})^2}, \qquad H_{\infty} \simeq H_s,
$$

or

$$ 
a(H_{\infty})=1-\sqrt{\frac{2\lambda_{eff}}{\pi W}}
\frac{H_s}{H_{\infty}}, \qquad H_{\infty} \gg H_s,
$$

while for $W/\lambda_{eff} \ll 1$ we have

$$
a(H_{\infty})=1-H_s/H_{\infty},
$$

for all values of $H_{\infty}$.

Using dependence (15), we can find distribution of the $z$-averaged magnetic 
field across a film width:

$$
\overline H_z(y)=
H_{\infty}+\frac{2}{c}\int_{-1}^{1} \frac{i(y')}{y-y'}dy'. \eqno(16)
$$

Fig. 9 shows the dependences $\overline H_z(y)$ for a film with 
$W/\lambda_{eff}=200$ and $a=0.6$ ($H_{\infty} \simeq 1.3H_s$), 
obtained numerically and by means of expression (15,16). 
It is seen that these dependences practically coincide across  
the entire width of a film, except for the near-edge regions and the 
boundary of the vortex-filled zone. The difference in 
the magnetic field value at $y=0.6$ should be attributed to the 
inaccuracy of approximation (more precisely, its first derivative at 
the boundary of the vortex-filled area). Dependence $\overline H_z(y)$ 
shown in the insert to Fig. 9 was obtained theoretically in \cite{3,5,9}.
It is seen that, as opposed to this analytical dependence, a non-zero 
magnetic field does exist outside the vortex-filled region also, 
and it is quite strong ($> 0.1 H_{\infty}$) for a film with the given 
parameters. Another distinguishing feature is the occurrence of a 
jump from zero to some finite value for the dependence 
$n(y)=\overline H_z(y)/\Phi_0$ at $y=a$. The reason of the vortex density
discontinuity is explained by a non-zero 
magnetic field in the region $(a, 1)$ and the condition of a 
magnetic field continuity at $y=a$.

Using Eq.(15), we can estimate the dependence of 
$\overline H_z^{edge}/H_{\infty}$ on the parameters of a film and 
an increasing external magnetic field (for thin films; see subsection 2.2)

$$
\frac{\overline H_z^{edge}}{H_{\infty}}=\frac{2\lambda_{eff}}{W}
\frac{1}{\sqrt{\beta}}\left(1+\frac{1}{\beta} \left(
2\alpha-\frac{\beta(1-a)}{2\sqrt{1+a^2/2}} \right) \right)
\frac{4}{1-a^2}. \eqno(17)
$$

It is nicely seen that in the limit $a \to 1$($H_{\infty} \to \infty$) 
the ratio $\overline H_z^{edge}/H_{\infty}$ tends to 1. 

Knowing the dependence $i(y)$, we can find the magnetization curves of 
superconducting films for different values of $W/\lambda_{eff}$. 
Fig.10 illustrates the obtained results. One can see that with a 
increasing parameter $W/\lambda_{eff}$ the magnetization curves 
become similar to the dependence $M(H)$ derived theoretically in 
\cite{5} for wide thin films. As $W/\lambda_{eff}$ decreases, the 
magnetization curves tend to the dependence which is valid for 
thin narrow films \cite{7,15}. Thus, even at $W/\lambda_{eff}=1$ 
the dependence $M(H)$ for a narrow 
film and the $M(H)$ obtained numerically by the use of Eq.(15) 
practically coincide. So, expression (15) allows to obtain 
magnetization curves for such superconductors at arbitrary ratio 
$W/\lambda_{eff}$. Note that the magnetization curves in 
this case, i.e., at any value of parameter $W/\lambda_{eff}$ lie between 
two curves - 1 and 3 as shown in Fig. 10 (in dimensionless units).

\section{Conclusion}

It is shown in the present paper that the Maxwell-London equation used so 
far only for thin films is also valid for samples of finite thickness.
This equation is shown to define the distribution of a thickness-averaged
vector potential and/or current density (in the mixed state case) 
across a sample width. For thin films the equation holds practically 
everywhere in a film, whereas in the thick film case its applicability 
is restricted only within a narrow bands near the edges 
$W/2-d/2 \leq |Y| \leq W/2$. 

An approximation expression is found, describing distribution of vector
potential $\overline A$ (or current density $\overline j(y)$) across the 
width of a film in the Meissner state, which applies to both thin 
and thick films. For thick films analytical expressions have been derived,
defining the value of the vector potential (local current density) at the
equator ($Y=1, Z=0$), side edges $(Y=1, Z=1)$, and also on the top and bottom 
surfaces ($Y, Z=\pm d/2$) of a sample. Besides, analytical approximation 
expressions have been found for the magnetic field at the equator and for 
the thickness-averaged edge magnetic field.

The vector potential distribution data were used to evaluate the field of the
the first vortex entry (barrier suupresion field) for superconductors of
such geometry. It's described by an universal expression (12) valid for
both thin and thick films. It is shown that besides surface defects the
layered structure of superconductor may result in a significant (up to 
$100 \%$) suppression of the vortex entry field. Thus, mutual influence
both surface defects and layered structure may lead to suppression of
$H_s$ by factor ten (and even greater).  

The study of a mixed state yields an interpolation expression for 
distribution of a sheet current density $i(y)$ in superconducting films 
without bulk pinning. This result allows to for the first time estimate 
the dependence on the magnetic field concentration 
$\gamma=\overline H_z^{edge}/H_{\infty}$ on the parameters of superconductor 
and external magnetic field $H_{\infty}$. 
Besides, these data were used to calculate the magnetization curves
for film superconductors at any values of parameter $Wd/\lambda^2$.

\section{Acknowledgements}

Authors are obliged to Prof. J.R.Clem, Dr. G.M. Maksimova 
for helpful discussions of the results obtained.
This work is supported by the Science Ministry of Russia (Project
No. 98-012), and in part Basic Foundation for Fundamental Research 
(Project No. 97-02-17437). Partial support of the International 
Center for Advanced Studies (INCAS) through Grant No. 99-2-03 is 
gratefully acknowledged.

\newpage

{\Large Figure captions}

fig.1

Geometry of the problem.

fig.2

Distribution of averaged vector potential, for different ratios
$W/\lambda_{eff}$: curves 1-5 for $W/\lambda_{eff}=1, 5, 10, 50, 200$,
respectively. Dotted lines are numerical results,
solid lines are approximation (8).

fig.3

Dependence of the parameter $\alpha$ on $W/\lambda_{eff}$. Circles are 
numerical results, solid curve 1 is the approximation (9).   

fig.4

Distribution of $\overline H_z$ inside the film for $W/\lambda_{eff}=200$. 
Dots are numerical result, solid line is expression obtained from 
approximation (8). Insert shows dependence $\overline H_z^{edge}
(W/\lambda_{eff})$: circles are numerical result, dotted line is 
the expression (10), solid line
is the fitting function $1+0.66\sqrt{W/\lambda_{eff}}$.

fig.5 

Distribution of averaged vector potential
at thick film ($d=10\lambda$) for various widths: $W=50\lambda$ (1), 
$W=100\lambda$ (2), $W=200\lambda$ (3). Dots are numerical result, 
solid lines are expression (8) with $\alpha=0.25$, $\beta=2\lambda^2/
\pi dW $. 

fig.6

Contour lines of the intensity of magnetic field (a,c) and current
density (b,d) of thick ($W=100\lambda$, $d=10\lambda$) film in applied
perpendicular magnetic field. The step for magnetic field is 
$0.41 H_{\infty}$, for current density is $0.1 j(1,1)$. 
Maximum values of magnetic field ($H=4.1H_{\infty}$) and current 
density ($j=1$) are reached at the corners of the film.

fig.7

Dependences $\overline H_z^{edge}$ (circles) and $H_z(1,0)$ (dots) on the
parameter $\sqrt{W/d}$. Solid line 1 is the fitting
function $1/3+1.03\sqrt{W/d}$, dotted line 2 is  the fitting function
$1/3+0.92\sqrt{W/d}$. Insert shows the dependences
of $\overline H_z^{edge}$(circles), $H_z(1,0)$(dots) and 
$H_z(1,1)$(stars) are shown on the film thickness for $W/d=5$. 

fig.8

Distribution of the sheet current $i(y)$ for film with parameter 
$W/\lambda_{eff}=200$ and for different values of $a$: 0.0 (1),
0.4 (2), 0.8 (3). Dotted lines are numerical results,
solid lines are expression (15).  

fig.9

Distribution of the averaged $z$-component of
magnetic field for $W/\lambda_{eff}=200$ and $a=0.6$. 
Solid line is obtained with help of expression (15,16), dotted line
is numerical result. Insert shows detailed distribution of the field;
dashed line is the function $H\sqrt{a^2-y^2}/\sqrt{1-y^2}$  
from \cite{5,9}.

fig.10

Magnetization curves of superconducting films for different 
ratio $W/\lambda_{eff}$: curve 1 for $W/\lambda_{eff}=\infty$,
curve 2 for $W/\lambda_{eff}=200$, curve 3 for $W/\lambda_{eff}=1$.
Curve 3 practically coincides with the magnetization curve for
narrow $W \ll \lambda_{eff}$ films \cite{15}. 

\end{document}